\documentclass[unnumsec,webpdf,contemporary,large]{main}%

\graphicspath{{Fig/}}

\theoremstyle{thmstyleone}%
\theoremstyle{thmstyletwo}%
\theoremstyle{thmstylethree}%

\DeclareUnicodeCharacter{2212}{-}

\begin{document}

\journaltitle{Preprint}
\DOI{DOI HERE}
\copyrightyear{2023}
\pubyear{2023}
\appnotes{Paper}

\firstpage{1}


\title[Vaxformer]{Vaxformer: Antigenicity-controlled Transformer for Vaccine Design Against SARS-CoV-2}

\author[1,$\dagger$,$\ast$]{Aryo Pradipta Gema\ORCID{0009-0007-1163-3531}}
\author[1,$\dagger$,$\ast$]{Micha{\l} Kobiela}
\author[1,$\dagger$,$\ast$]{Achille Fraisse}
\author[1,$\ddagger$]{Ajitha Rajan\ORCID{0000-0003-3765-3075}}
\author[1,2,3,$\ddagger$]{Diego A. Oyarz\'un\ORCID{0000-0002-0381-5278}}
\author[1,4,5,$\ddagger$,$\ast$]{Javier Antonio Alfaro\ORCID{0000-0002-5553-6991}}

\authormark{Gema et al.}

\address[1]{\orgdiv{School of Informatics}, \orgname{University of Edinburgh}, \orgaddress{\state{Edinburgh}, \country{United Kingdom}}}
\address[2]{\orgdiv{School of Biological Sciences}, \orgname{University of Edinburgh}, \orgaddress{\state{Edinburgh}, \country{United Kingdom}}}
\address[3]{\orgname{The Alan Turing Institute}, \orgaddress{\state{London}, \country{United Kingdom}}}
\address[4]{\orgdiv{International Centre for Cancer Vaccine Science}, \orgname{University of Gda\'nsk}, \orgaddress{\state{Gda\'nsk}, \country{Poland}}}
\address[5]{\orgdiv{Department of Biochemistry and Microbiology}, \orgname{University of Victoria}, \orgaddress{\state{British Columbia}, \country{Canada}}}

\corresp[$\dagger$]{Joint first author.}
\corresp[$\ddagger$]{Joint senior authorship.}
\corresp[$\ast$]{Corresponding authors: \href{aryo.gema@ed.ac.uk}{aryo.gema@ed.ac.uk}, \href{s1870794@ed.ac.uk}{s1870794@ed.ac.uk}, \href{s1898908@ed.ac.uk}{s1898908@ed.ac.uk}, \href{javier.alfaro@proteogenomics.ca}{javier.alfaro@proteogenomics.ca}}




\abstract{
\textbf{Motivation:} The SARS-CoV-2 pandemic has emphasised the importance of developing a universal vaccine that can protect against current and future variants of the virus. \\
\textbf{Results:} The present study proposes a novel conditional protein Language Model architecture, called Vaxformer, which is designed to produce natural-looking antigenicity-controlled SARS-CoV-2 spike proteins. We evaluate the generated protein sequences of the Vaxformer model using DDGun protein stability measure, netMHCpan antigenicity score, and a structure fidelity score with AlphaFold to gauge its viability for vaccine development.
Our results show that Vaxformer outperforms the existing state-of-the-art Conditional Variational Autoencoder model to generate antigenicity-controlled SARS-CoV-2 spike proteins. 
These findings suggest promising opportunities for conditional Transformer models to expand our understanding of vaccine design and their role in mitigating global health challenges. \\
\textbf{Availability and Implementation:} The code used in this study is available at \href{https://github.com/aryopg/vaxformer}{https://github.com/aryopg/vaxformer}.
}


\maketitle

\section{Introduction}

The severe acute respiratory syndrome coronavirus 2 (SARS-CoV-2) virus is a pathogen of the coronavirus family. Since its discovery in 2019 \cite{wang_novel_2020}, it caused major health outbreaks all over the world.
A range of vaccines eventually enabled communities to ease their restrictive measures.

The SARS-CoV-2 virus eventually evolved through random mutations creating variants with the potential to escape the immunity generated by our current vaccines \cite{flemming_omicron_2022}. Currently, more than 30 vaccines have been used \cite{noauthor_covid-19_nodate}, mostly based on a specific variant of the virus. A more efficient approach might be to develop synthetic vaccines that are 'universal', meaning they can guard against a breadth of current and potential future strains of the virus. Practically speaking, this means studying the current sequence diversity of the virus in the context of its immune visibility in the context of a specific human population. 

mRNA vaccines were the first approved vaccines against SARS-CoV-2.
Their mechanism of action involves introducing mRNA, which codes for a viral protein, directly into cells where it can be read and utilised. The viral fragment encoded is presented to the immune system to generate an immune response,
training our body to recognise and eliminate the virus \cite{dolginTangledHistoryMRNA2021}.  The parts of a protein that are recognized by the immune system are called antigens. Antigens bind to specific receptors on B-cells or are presented to T-cells via major histocompatibility complex (MHC) molecules, to trigger an immune response.
The optimal vaccine component is therefore a protein located on the virus's surface, facilitating immune recognition through both B-Cell and T-Cell immunity.
In SARS-CoV-2, the target protein was the \emph{spike protein}
\cite{huangStructuralFunctionalProperties2020} and particular strains of the \emph{spike protein} were used.

One way vaccines against SARS-Cov-2 could be improved is by studying the known evolutionary breadth of the virus to produce synthetic spike protein sequences that cover more of the antigenic plasticity of existing SARS-CoV-2 spike proteins, and that could ideally safeguard against future variants. Such a 'universal' vaccine would provide protection more broadly against SARS-CoV-2, its past and current geographic variation. So, tools that can create synthetic Spike protein sequences that capture more of this diversity, and with respect to the immunity of a target human population are needed to improve the efficacy of future vaccine efforts. A key feature the sequence must have is antigenicity, its capacity to replicate those parts of the virus that can be recognised by the immune
system. 
In vaccine development, high antigenicity to both T-Cell and B-Cell responses is desirable to prime the immune system. Conversely, low antigenicity is advantageous for applications such as drug design, where persistence is desired \cite{shakhnovichImmunogenicityClinicalPractice2020}. Yet, the synthetic sequences produced must also resemble the natural sequences of the virus, this is indeed a requirement to ensure B-Cell functionality. With the amount of data we now have on SARS-CoV-2 and its variants \cite{shuGISAIDGlobalInitiative2017}, machine learning could prove to be the solution to generate protein sequences with such constraints.

A recent study utilised Variational Autoencoder (VAE) to generate structurally valid and stable Covid spike proteins~\cite{phillipsGeneratingImmuneawareSARSCoV22022}. The proteins generated by VAE, however, were only marginally more stable and slightly less structurally valid than those generated by the n-gram baseline model. One explanation for this outcome is that VAE uses Euclidean distance as a similarity metric which can be insufficient for tasks with complex data distributions \cite{dosovitskiy2016generating, larsen2016autoencoding}. In the domain of image generation, a tiny translation of the image can result in low similarity while humans barely can notice the difference \cite{larsen2016autoencoding}. Similarly, different pointwise mutations have different effects on the three-dimensional structure of the protein and its stability. Simple pointwise metrics such as Euclidean distance likely fail to capture such dependencies.

A potentially more suitable similarity metric would be the Root-mean-square deviation (RMSD) of atomic positions of the residues in three-dimensional models of the protein rather than distance based on unfolded sequences of amino acids (AAs).
However, the computational feasibility of this approach is hindered by the extended duration required for protein folding.
Therefore, models that can learn more complex similarity measures during training are promising solutions to this problem.

Additionally, the VAE solution~\cite{phillipsGeneratingImmuneawareSARSCoV22022} is limited in terms of its scalability. The VAE model depends heavily on the choice of hyperparameters, making it inherently hard to be fine-tuned to new tasks, i.e. new vaccine design tasks. Therefore, there is a need to develop a foundational model that can overcome these limitations.

In recent years, several studies have applied Language Models (LMs) to protein design, with the aim of developing a foundational model that can comprehend the underlying patterns of protein design.
One of the earliest protein LMs utilised a Long short-term memory~\citep[LSTM;][]{hochreiterLongShortTermMemory1997} architecture to generate single-domain antibodies~\cite{shinProteinDesignVariant2021}.
There has been a growing interest in applying state-of-the-art NLP architectures to the field of protein design.
A recent study proposed ProtGPT2, a model to generate novel protein sequences~\cite{ferruzProtGPT2DeepUnsupervised2022}, which is based on the GPT2 architecture~\cite{radfordLanguageModelsAre2019} and contains 738 million parameters.
ProtGPT2 was trained to predict the next oligomer in the sequence using the UniRef50 dataset which comprises 44.9 million training sequences and 4.9 million validation sequences.

Another recently proposed protein LM is ProGen~\cite{madaniLargeLanguageModels2023} that is trained to generate protein sequences that are conditioned to a control tag, such as protein family, biological process and molecular function.
ProGen is based on the CTRL architecture~\cite{keskarCTRLConditionalTransformer2019}, which has 1.2 billion parameters.
For the training of ProGen, 281 million non-redundant universal protein sequences and their associated properties were collected from various sources, such as UniParc~\cite{leinonenUniProtArchive2004}, UniprotKB~\cite{bairochUniversalProteinResource2004}, Pfam~\cite{finnPfamProteinFamilies2014}, and NCBI taxonomic information~\cite{federhenNCBITaxonomyDatabase2012}.

Unfortunately, both ProtGPT2 and ProGen require considerable computing resources, making it challenging to even fine-tune them for the spike protein design task.
On top of that, there is an absence of studies exploring the use of LMs in designing spike proteins thus far, specifically controllable vaccine design based on antigenicity scores.

In this study, we attempt to fill in this gap by proposing Vaxformer, a conditional spike protein LM architecture which is designed to produce antigenicity-controlled SARS-CoV-2 vaccines.
The quality of hypothetical proteins generated by Vaxformer is mainly evaluated by three metrics: protein stability computed by the DDGun \cite{montanucci2019ddgun} tool, similarity to reference spike protein measured by RMSD of folded models by AlphaFold 2 \cite{jumper2021highly} and netMHCpan~\cite{reynissonNetMHCpan4NetMHCIIpan4Improved2020} antigenicity scores.
The Vaxformer model's ability to produce spike proteins that are both stable and structurally sound while being conditioned on antigenicity scores can make a noteworthy contribution to the study of vaccine design against SARS-CoV-2.

\section{Materials and Methods}

\subsection{Dataset and Task}

The dataset of spike Covid proteins was obtained from GISAID \cite{khare2021gisaid}. Each protein in the dataset is represented as a sequence of amino acids (AAs). There are 20 possible AAs in each position of the sequence. The proteins were first aligned using Multiple Sequence Alignment (MSA) with the MUSCLE tool \cite{edgar2004muscle}. Then we used one-hot encoding to convert each of the aligned sequences to binary data. Finally, the dataset was split into 65,027 training samples (95\%), 2,000 validation samples (2.5\%), and 2,000 test samples (2.5\%). The antigenicity score of every sequence in the dataset was measured using netMHCpan to get the number of ``hits" of each sequence (a ``hit" being a 9-aa peptide from the sequence that is likely to be presented to the immune system). Then, the sequences were assigned with an antigenicity score (0, 1 or 2) using the first and third quantiles of the number of hits in the dataset.
The task is to generate a sequence of amino acids which is a stable and structurally valid spike protein conditioned on antigenicity score.

\subsection{Models}

\subsubsection{Naive Bayes Baseline}

We developed a Naive Bayes \cite{bishop2006pattern} model as a baseline for this study. The Naive Bayes model is trained to calculate the most likely AA in a given position of a sequence.
The Naive Bayes model is not trained to be conditioned to antigenicity scores.
Naive Bayes models each position independently.
The frequency of each residue is calculated for a given position and normalised, which can be interpreted as probabilities.
By sampling a categorical distribution with those probabilities we generated synthetic sequences.
While this approach is computationally efficient, the model cannot capture which mutations are likely to occur together.

\subsubsection{Long Short Term Memory}

\begin{figure}
    \centering
    \includegraphics[width=0.7\linewidth]{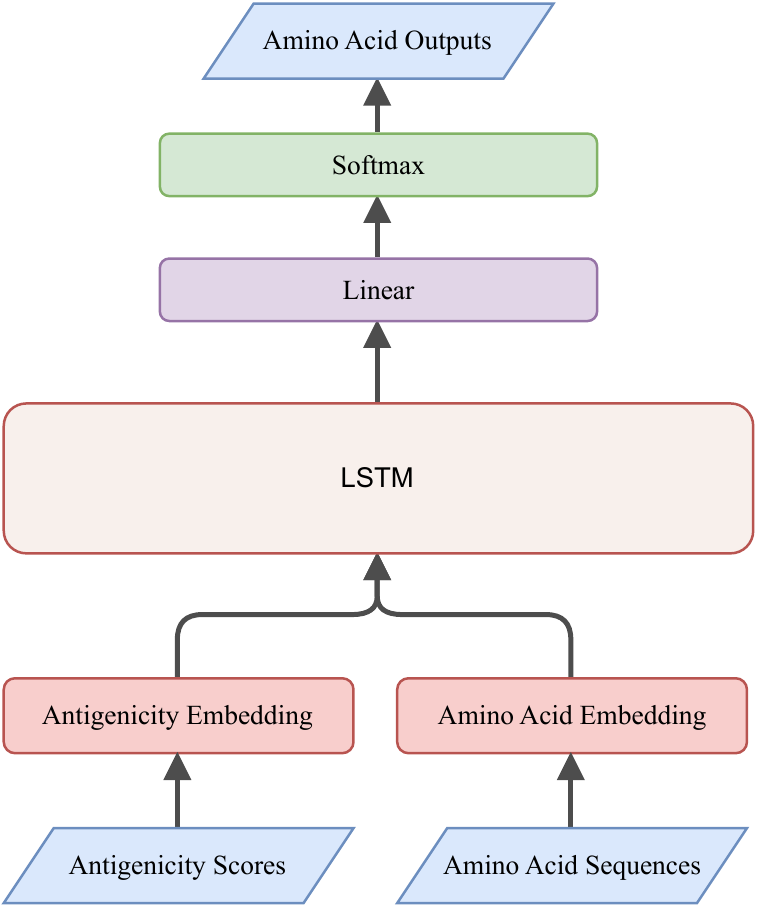}
    \caption{LSTM architecture. The LSTM architecture employs two separate embeddings to map AA tokens and antigenicity scores into their respective vectors, which are concatenated and passed to the LSTM model. The LSTM layer produces a latent representation which is ingested by a softmax linear layer to predict the subsequent AA token.}
    \label{fig:lstm}
\end{figure}

The LSTM model is trained to generate sequences from left to right with prior AAs and their antigenicity scores as context, as shown in~\autoref{fig:lstm}.
The LSTM model takes aligned protein sequences with a fixed length of 1,299 which contains AA or shift characters (``-").
The LSTM model also accepts the corresponding batch of antigenicity scores (0, 1, or 2) which helps condition the LSTM prediction.
Both antigenicity scores and the AA sequences are mapped to a high-dimensional vector by the embedding layer.
After both respective embedding layers, both antigenicity and AA vectors are concatenated and ingested by the LSTM. This can be formally defined as:
\begin{equation}
    \begin{aligned}
    x_t & = a_t \parallel s_t \\
    i_t & =\sigma\left(W_i x_t + U_i h_{t-1} + b_i\right) \\
    f_t & =\sigma\left(W_f x_t + U_f h_{t-1} + b_f\right) \\
    \tilde{c}_t & =\tanh \left(W_c x_t + U_c h_{t-1} + b_c\right) \\
    c_t & =\left(i_t \odot \tilde{c}_t + f_t \odot c_{t-1}\right) \\
    o_t & =\sigma\left(W_o x_t + U_o h_{t-1} + b_o \right) \\
    h_t & = \tanh \left(C_t\right) * o_t
    \end{aligned}
\end{equation}
where $a_t$ and $s_t$ denote the amino acid and the antigenicity embedding vector of time step $t$, $\parallel$ indicates a concatenation along the last axis, and $\odot$ denotes the Hadamard product. $\sigma$ and $\tanh$ denote a sigmoid and a hyperbolic tangent activation function, respectively. $W_i$, $U_i$, $W_f$, $U_f$, $W_c$, $U_c$, $W_o$, and $U_o$ denote the input, forget, memory, and output gate matrices of the LSTM, while $b_i$, $b_f$, $b_c$, and $b_o$ denote the input, forget, memory, and output gate biases of the LSTM. $o_t$, $h_t$ and $c_t$ denote the output, hidden state, and cell state of time step $t$. We employed a fully-connected layer with softmax activation to predict the most probable AA for a specific timestamp.

The best performing LSTM uses 2 LSTM layers, a hidden dimension of 128, and an embedding dimension of 128.
The LSTM is trained with Adam optimiser~\cite{kingmaAdamMethodStochastic2017} with a learning rate of $1 \times 10^{-4}$ for 150 epochs.

\subsubsection{Vaxformer}

\begin{figure}
    \centering
    \includegraphics[width=\linewidth]{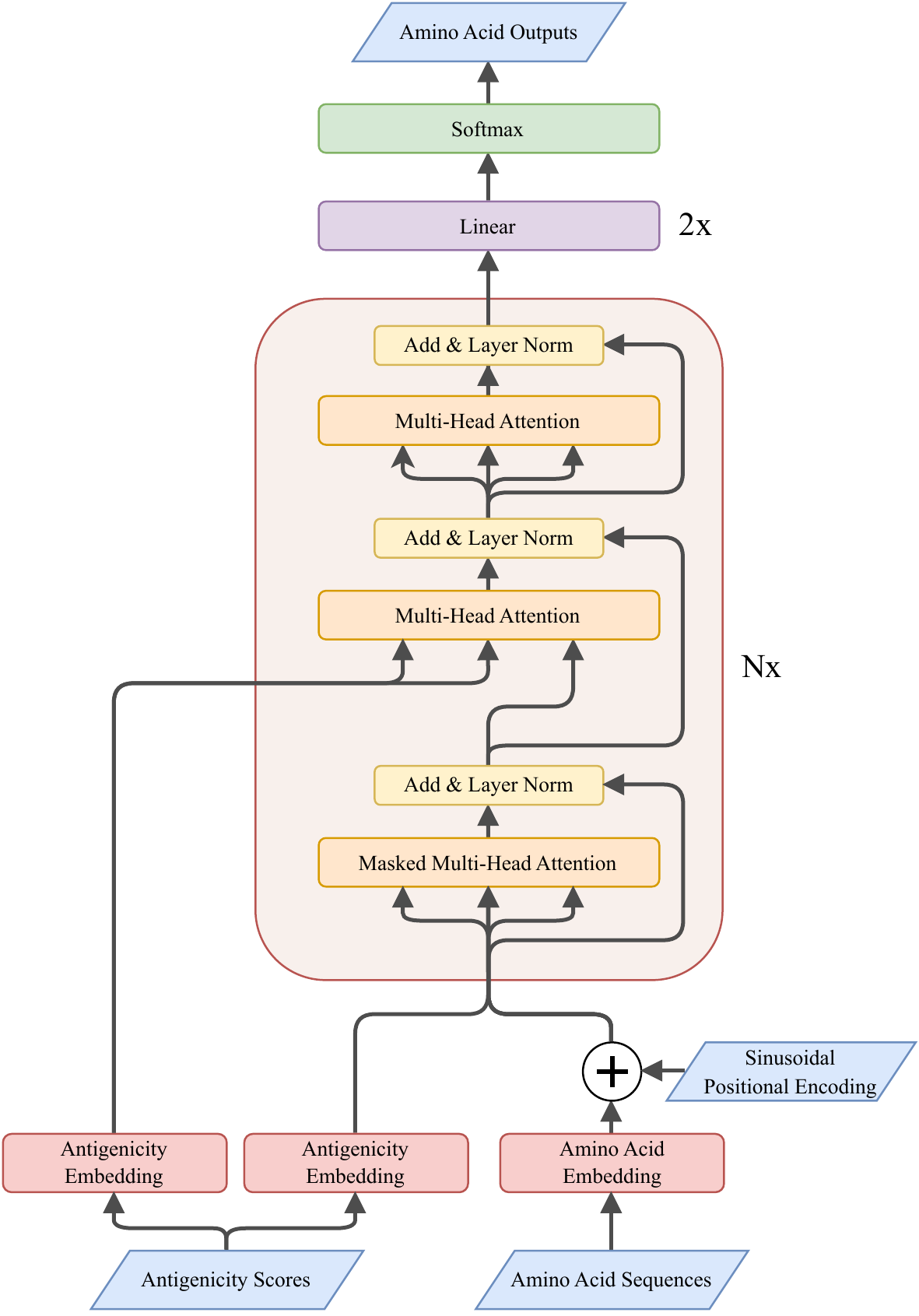}
    \caption{Vaxformer architecture. Alongside the AA tokens embedding, the Vaxformer architecture employs two separate embeddings to map the Antigenicity scores; one embedding will be concatenated with the AA vectors, one embedding will act as the key-value of the multi-head attention calculation. From this combination of inputs, the Vaxformer produces a latent representation which is ingested by a softmax linear layer to predict the subsequent AA token.}
    \label{fig:vaxformer}
\end{figure}

Vaxformer is a Transformer model~\cite{vaswaniAttentionAllYou2017} that is trained in an autoregressive manner, generating sequences from left to right. Vaxformer generally follows the design of the original Transformer model.
Vaxformer takes aligned protein sequences with fix length of 1,299 which contains AA or shift characters.
The protein sequences will be processed by the Amino Acid Embedding and injected by the Sinusoidal Fixed Positional Embedding, which aims to introduce information about the order of the tokens in the input sequence. We use sine and cosine functions similar to the original Transformer:

\begin{equation}
\begin{aligned}
P E_{(p o s, 2 i)} & =\sin \left(p o s / 10000^{2 i / d_{\mathrm{model}}}\right) \\
P E_{(p o s, 2 i+1)} & =\cos \left(p o s / 10000^{2 i / d_{\mathrm{model}}}\right)
\end{aligned}
\end{equation}

Alongside the AA sequences, Vaxformer also takes the corresponding batch of antigenicity scores (0, 1, or 2) which allows Vaxformer to learn to generate spike protein sequences conditioned to those antigenicity scores.
As presented in~\autoref{fig:vaxformer}, the antigenicity scores are processed by two separate embeddings. One embedding will be concatenated to the AA sequence embedding, while the other will act as a key and value of the multi-head attention calculation.
This can be formally defined as:
\begin{equation}
\operatorname{Attention}(Q, C)=\operatorname{softmax}\left(\frac{Q C^T}{\sqrt{d_C}}\right) C
\end{equation}
where $Q$ denotes the amino acid sequences, $C$ denotes the antigenicity scores, and $d_C$ denotes the hidden dimension of the input. Notice that the antigenicity scores act as the key and value of the multi-head attention computation, forcing the model to condition its latent representation to the antigenicity scores. In our experiment, we used two separate antigenicity embeddings (to condition the AA sequences and to be the key-value of the multi-head attention computation), but theoretically, both embeddings can share the same weight.
Vaxformer is trained to minimise the cross-entropy loss by comparing the output probability of AAs to the real sequences. 

The best-performing Vaxformer uses a hidden dimension of 200, 4 attention heads, 3 layers, dropout rate of 0.5 on the embedding layer. Vaxformer is trained with Adam optimiser~\cite{kingmaAdamMethodStochastic2017} with a learning rate of $1 \times 10^{-4}$ for 100 epochs.

\subsection{Evaluation methods}\label{subsec:eval_methods}

The evaluation process comprises two steps. The first step is to evaluate the Protein LMs using AA-level perplexity. Perplexity evaluates the efficacy of a LM in predicting the probability of the subsequent AA in the entire sequence. It measures the uncertainty of the model when attempting to predict the next AA, and a lower perplexity score implies better predictive power.
We identify the best-performing LMs per model family (i.e. LSTM and Vaxformer).

Once the LMs are selected, we evaluate the quality of the generated proteins in comparison to the random mutation and Naive Bayes baseline, as well as the state-of-the-art VAE model~\cite{phillipsGeneratingImmuneawareSARSCoV22022}. The four main ways of evaluating the quality of generated protein are:

\begin{itemize}
    \item \textbf{Protein stability computed by DDGun.} Stability in this context refers to the net balance of forces which indicates whether a mutated protein will keep its native shape or it will unfold or extend \cite{gromiha2010protein}. We identify the most similar natural protein to the generated one based on sequence alignment score using BLOSUM62 similarity matrix. DDGun is then used to calculate the difference in the free energy of unfolding between the natural and generated protein. A lower score indicates greater stability. 
    \item \textbf{RMSD of folded protein by the AlphaFold 2 and Covid reference protein.} We first use AlphaFold to compute the three-dimensional structure of the generated protein. Using PyMOL \cite{delano2002pymol} software we align the reference Wuhan protein with folded generated protein and compute the RMSD.
    \item \textbf{Protein antigenicity evaluation by netMHCpan.} We use netMHCpan to predict the efficiency of a protein to trigger an immune response. This involves predicting the binding of ~9 AA peptides to the MHC complex
    using a feed-forward neural network. Antigenicity scores are calculated by determining the number of compatible rolling windows of 9-mers in a sequence compared to the rest of the dataset. The analysis is conducted on 12 MHC versions common in human populations, and we compare the antigenicity score distributions of the generated sequences to the desired distributions.
    \item \textbf{PCA visualisation.} We mapped the natural sequences into two-dimensional latent space using the Principal Component Analysis (PCA). We use the same transformation to map generated proteins which allows us to qualitatively compare their positions relative to natural proteins. This serves as a qualitative evaluation of the generated proteins to identify whether the generated proteins are distributed similarly as the natural proteins.
\end{itemize}

\section{Results}
\label{sec:results}

\subsection{Language Model Perplexity}
\begin{table*}[htbp]
    \centering
    \caption{Train and validation AA-level perplexities of the experimented LMs. Lower perplexity indicates a better model. The lower bound of perplexity is 1. The \textbf{boldface row} indicates the best-performing model from each category. ''-small'', ''-base'', and ''-large'' models differ in terms of their embedding dimension, hidden dimension, and the number of layers. While the rest of the hyperparameters remain the same.}
    \begin{tabular}{lccccc}
        \toprule
        Model & Emb. dim & Hidden dim & Num. layers & Train Perplexity  & Val. Perplexity  \\ \midrule
        LSTM-small & 32 & 64 & 2 & 1.019            & 1.021           \\
        \textbf{LSTM-base} & \textbf{128} & \textbf{128} & \textbf{2} & \textbf{1.016}   & \textbf{1.018}  \\
        LSTM-large & 200 & 200 & 4 & 1.016            & 1.019           \\ \midrule
        Vaxformer-small & 64 & 64 & 1 & 1.034            & 1.128           \\
        Vaxformer-base & 128 & 128 & 2 & 1.014            & 1.043           \\
        \textbf{Vaxformer-large} & \textbf{200} & \textbf{200} & \textbf{3} & \textbf{1.013}   & \textbf{1.014}  \\
        \botrule
    \end{tabular}
    \label{table:perplexity}
\end{table*}

\autoref{table:perplexity} exhibits the perplexity achieved by all trained LMs. Vaxformer-large achieved the lowest validation perplexity of 1.014, whereas, among the LSTM variants, LSTM-base achieved the lowest validation perplexity of 1.018. We can observe that in Vaxformer exists a direct correlation between the model size and its performance, as the largest model yielded the lowest perplexity and the smallest model yielded the highest. We chose LSTM-base and Vaxformer-large to generate sequences for the evaluation.

\subsection{Generated Sequences}

\begin{table}
    \centering
    \caption{Total number of distinct and novel sequences arising from 2,000 generated sequences from Vaxformer, LSTM, and Naive Bayes as well as 50,000 generated sequences from VAE.}
    \begin{tabular}{lccc}
    \toprule
    Model              & \# Sample & \# Distinct & \# Novel \\ \midrule
    Naive Bayes        & 2,000     & 2,000       & 1,999    \\ \midrule
    VAE (low)    & 50,000    & 932         & 875      \\
    VAE (medium) & 50,000    & 1,308       & 1,110    \\
    VAE (high)   & 50,000    & 1,298       & 1,082    \\ \midrule
    LSTM-base (low)    & 2,000     & 1,015       & 933      \\
    LSTM-base (medium) & 2,000     & 891         & 780      \\
    LSTM-base (high)   & 2,000     & 1,283       & 1,243    \\ \midrule
    Vaxformer-large (low)    & 2,000     & 1,770       & 1,770    \\
    Vaxformer-large (medium) & 2,000     & 1,981       & 1,981    \\
    Vaxformer-large (high)   & 2,000     & 1,871       & 1,871    \\
    \botrule
    \end{tabular}
\label{table:generation_stats}
\end{table}

We proceed to generate hypothetical protein sequences using the Naive Bayes baseline model, LSTM-base, and Vaxformer-large. Each model may generate the same protein sequences multiple times. It is imperative to differentiate the distinct proteins and eliminate those already present in the training data to obtain the \textit{de novo} sequences. The number of generated sequences, including the distinct and \textit{de novo} ones, are listed in~\autoref{table:generation_stats}.

\begin{table}
\centering
\caption{Mean values for RMSD w.r.t. the Wuhan AlphaFold and $\Delta \Delta G$. The \textbf{boldface} indicates the best-performing result from each metric.}
    \begin{tabular}{lcccc}
    \toprule
    Model           & RMSD (\AA)   & $\Delta \Delta G$ (KCAL/MOL)      \\ \midrule
    Random Mutation~\cite{phillipsGeneratingImmuneawareSARSCoV22022} & \textbf{0.32 $\pm$ 0.23} & -2.51 $\pm$ 0.31                  \\
    Naive Bayes     &   $0.59 \pm 0.23$                  &  -0.5 $\pm$ 0.30                            \\
    VAE~\cite{phillipsGeneratingImmuneawareSARSCoV22022}             & 0.48 $\pm$ 0.28  & −5.17 $\pm$ 0.51                  \\
    LSTM-base       & 0.51 $ \pm 0.17 $                 &  -4.95 $\pm$ 1.09                            \\
    Vaxformer-large & 0.67 $\pm$ 0.31                  &  \textbf{-5.45 $\pm$ 0.72}                            \\
    \botrule
    \end{tabular}
\label{table:structural_stats}
\end{table}

\begin{table*}
\centering
\caption{Average one-tailed T-statistics and U-statistics of the antigenicity score distributions across 10 iterations of 750 randomly sampled model-generated sequences. All test results are statistically significant (p-value < 0.01) with negligible standard deviations of T-statistics and U-statistics. The higher the T-statistic or U-statistic value, the greater the difference between the two distributions, where the left-hand distribution is larger than the right-hand (e.g. Medium-Low indicates Medium as the left-hand distribution and Low as the right-hand distribution). The \textbf{boldface} row indicates the largest T-statistics and U-statistics.}
    \begin{tabular}{l ccc ccc}
    \toprule
    \multirow{2}{*}{Model} & \multicolumn{3}{c}{T-statistics} & \multicolumn{3}{c}{U-statistics}    \\
                           & Medium-Low     & High-Medium     & High-Low     & Medium-Low      & High-Medium      & High-Low      \\  \midrule
    VAE                    & 15.22   & 27.68   & 39.17   & $3.99 \times 10^5$ & $4.73 \times 10^5$ & $5.23 \times 10^5$ \\
    LSTM-base              & 14.23   & 16.83   & 29.75   & $4.45 \times 10^5$ & $4.37 \times 10^5$ & $5.16 \times 10^5$ \\
    \textbf{Vaxformer-large}        & \textbf{26.30}   & \textbf{38.02}   & \textbf{56.62}   & $\bm{4.75 \times 10^5}$ & $\bm{5.21 \times 10^5}$ & $\bm{5.56 \times 10^5}$ \\
    \botrule
    \end{tabular}
\label{table:t_test}
\end{table*}

\subsection{Protein stability}
Using the DDGun tool, we selected the ten most stable proteins for each of the models.
\autoref{table:structural_stats} presents the $\Delta\Delta G$ for each of the models. The Vaxformer achieved state-of-the-art performance and produced the most stable proteins. Conversely, the baselines (random mutator and Naive Bayes) were the least performant. LSTM also achieved good protein stability, however, both VAE and Vaxformer proved to be superior in that evaluation metric. 

\subsection{Structural similarity}

\begin{figure*}[htbp]
    \centering
    \begin{minipage}[t]{0.3\linewidth}
      \centering
      \includegraphics[width=\linewidth]{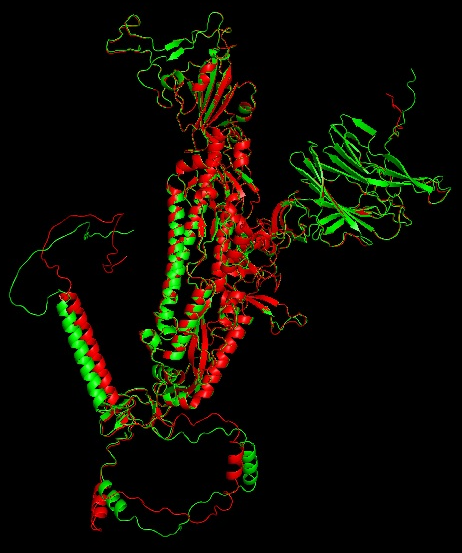}
      (a) Reference + Vaxformer.
      \label{fig:vae_af}
    \end{minipage}
    \begin{minipage}[t]{0.3\linewidth}
      \centering
      \includegraphics[width=\linewidth]{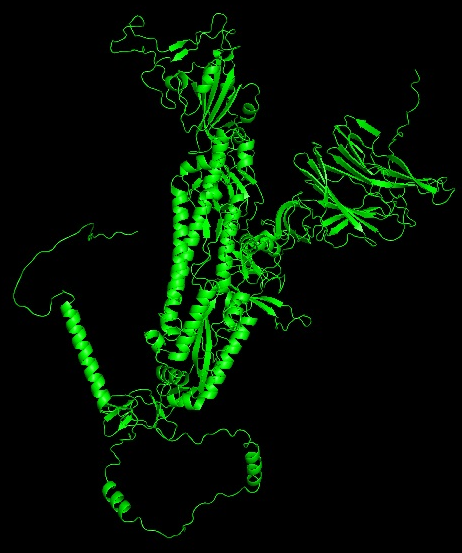}
      (b) Reference protein.
      \label{fig:lstm_af}
    \end{minipage}
    \begin{minipage}[t]{0.3\linewidth}
      \centering
      \includegraphics[width=\linewidth]{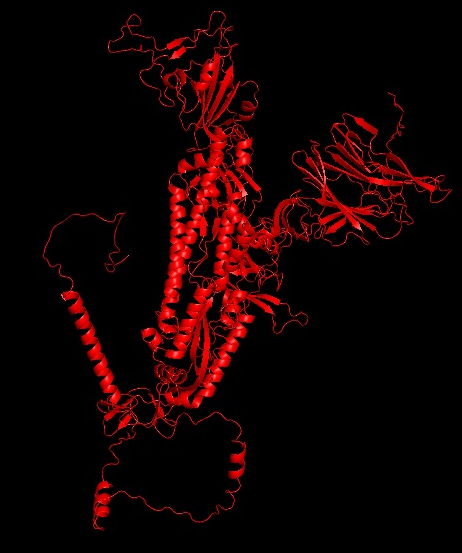}
      (c) Vaxformer generated protein.
      \label{fig:vaxformer_af}
    \end{minipage}
    \caption{Qualitative comparison of the reference and Vaxformer generated protein structures. Note the clear visual similarity. Alphafold and PyMOL were used to approximate and visualise the three-dimensional structures of the proteins.}
    \label{fig:alphafold}
\end{figure*}

\autoref{table:structural_stats} presents the RMSD values between the ten most stable proteins generated by each model and the Wuhan reference protein. Each mutation introduces some structural deviation, but all models produced structurally valid proteins, as shown by the average RMSD of natural proteins ($0.65 \pm 0.33 \AA$). The Random Mutation baseline had the lowest RMSD, while Vaxformer had the highest but was still comparable to the deviation of natural proteins. Additionally, \autoref{fig:alphafold} qualitatively compares the Vaxformer-generated protein with the reference Wuhan protein, demonstrating good conservation of the helices and high similarity between the two.

The hamming distance between the generated protein to the closest natural protein averaged over 10 most stable generated proteins was $10.73 \pm 1.889$ for the Vaxformer and $9.4 \pm 2.07$ for LSTM. This shows that the models managed to produce structurally valid and stable proteins which are notably divergent from the naturally occurring proteins. The confidence in AlphaFold predictions was measured using the averaged pLDDT metric (which has a range of 0-100 and higher means more confidence) which resulted in 77.26, 77.22 for Vaxformer and LSTM, respectively. This indicates that AlphaFold predictions were plausible.

\subsection{Antigenicity distributions}

\begin{figure*}[htbp]
    \centering
    \begin{minipage}[t]{0.325\linewidth}
      \centering
      \includegraphics[scale=0.225]{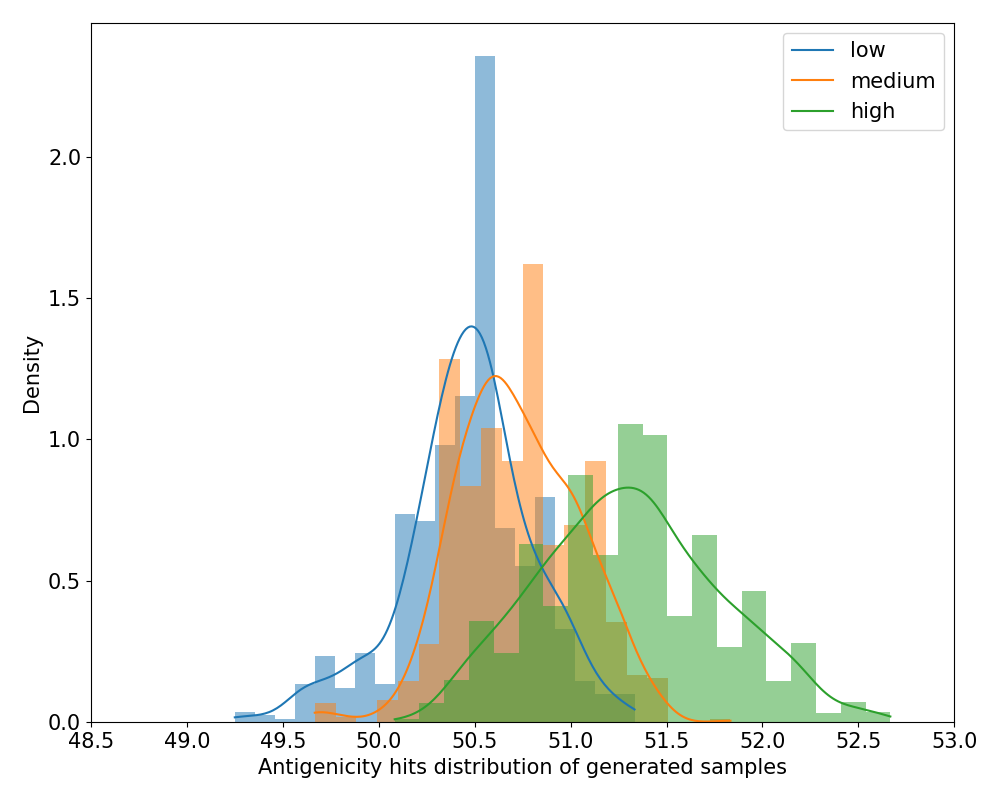}
      (a) VAE.
      \label{fig:vae_as}
    \end{minipage}
    \begin{minipage}[t]{0.325\linewidth}
      \centering
      \includegraphics[scale=0.225]{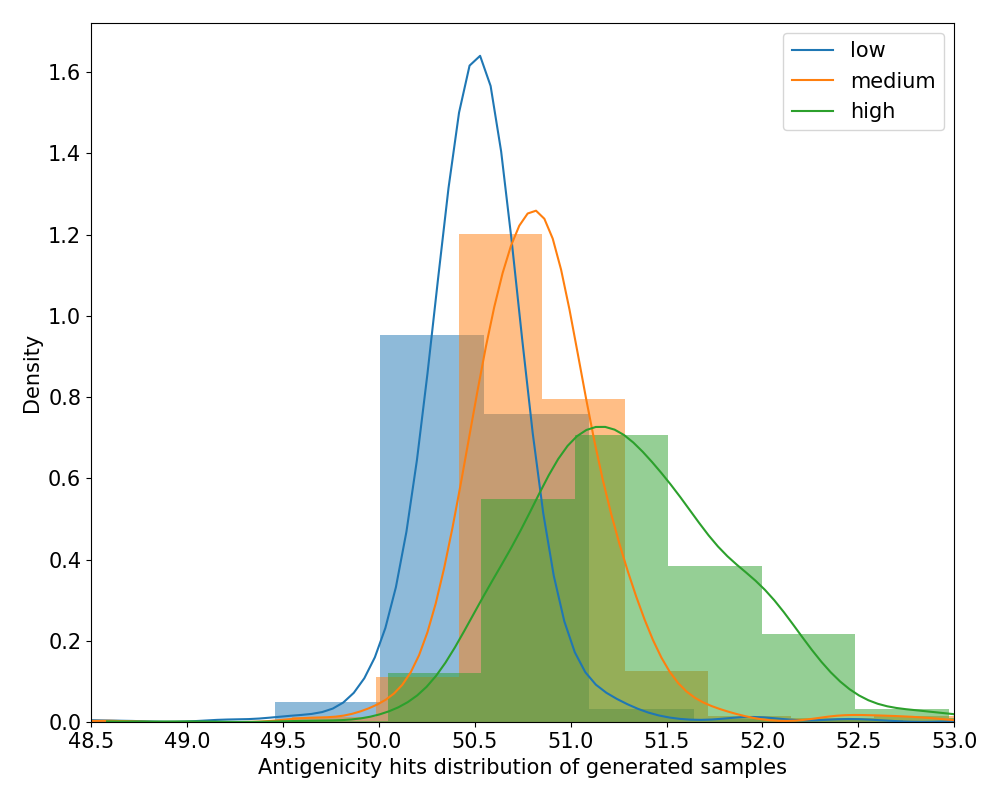}
      (b) LSTM.
      \label{fig:lstm_as}
    \end{minipage}
    \begin{minipage}[t]{0.325\linewidth}
      \centering
      \includegraphics[scale=0.225]{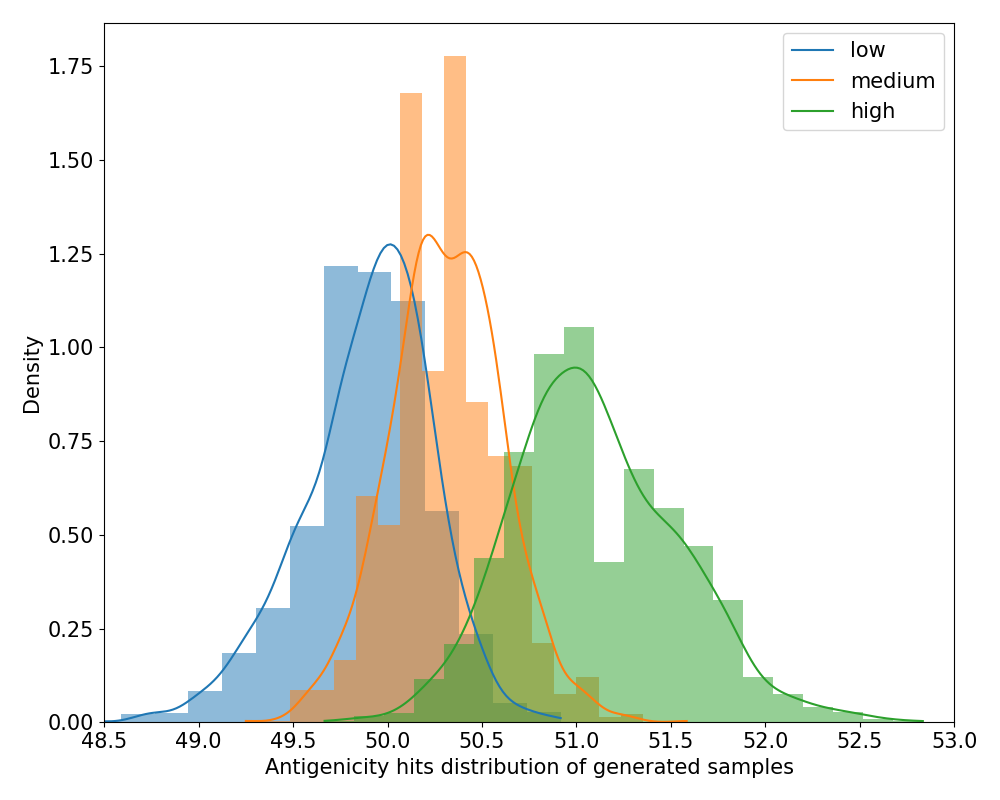}
      (c) Vaxformer.
      \label{fig:vaxformer_as}
    \end{minipage}
    \caption{Antigenicity score distributions of sequences generated by (a) VAE~\cite{phillipsGeneratingImmuneawareSARSCoV22022}, (b) LSTM, and (c) Vaxformer. Note that there are dubious sequences generated by LSTM with antigenicity lower than 48.5 due to excessive alignment gap tokens (``-").}
    \label{fig:AS_distributions}
\end{figure*}

We assessed the netMHCpan number of hits for sequences generated by VAE~\cite{phillipsGeneratingImmuneawareSARSCoV22022}, LSTM-based, and Vaxformer-large models.
The distributions of the antigenicity scores for the generated sequences are displayed in \autoref{fig:AS_distributions}, using a colour-coding scheme where blue, orange, and green represent low, medium, and high antigenicity scores, respectively.

To evaluate the separation between the antigenicity-modulated sequences, we conducted one-tailed T-tests and U-tests between the distributions of low-, medium-, and high-antigenicity sequences.
Because the goal of the project is to develop a model that can generate antigenicity-modulated sequences, it is important that the distributions of the generated sequences are well separated and ordered according to the antigenicity score (i.e. low $<$ medium $<$ high).
Both T-tests and U-tests produce a test statistic, the T-statistic for T-tests and U-statistic for U-tests, respectively, and a p-value. The p-value indicates the level of evidence against the null hypothesis, with smaller p-values indicating stronger evidence against the null hypothesis.
The T-statistic measures the difference between the means of the two groups relative to the variation within each group. A larger absolute value of the T-statistic indicates a greater likelihood that the means of the two groups are truly different. On the other hand, the U-statistic measures the difference between the medians of the two groups or the degree to which one group tends to have higher values than the other. A larger absolute value of the U-statistic indicates a greater likelihood that the medians of the two groups are truly different.

Our results, presented in \autoref{table:t_test}, indicate that Vaxformer-generated sequences exhibit superior antigenicity modulation compared to LSTM- and VAE-generated sequences. The T-statistics between the high- and low-antigenicity scores of Vaxformer-generated sequences (56.62) was significantly higher than that of VAE (39.17) and LSTM (29.75) sequences. A similar trend can be observed in the U-test results. The U-statistics between the high- and low-antigenicity scores of Vaxformer-generated sequences ($5.56 \times 10^5$) was higher than that of VAE ($5.23 \times 10^5$) and LSTM ($5.16 \times 10^5$) sequences.

\subsection{PCA Visualisation}

\begin{figure*}[htbp]
    \centering
    \begin{minipage}[t]{0.4\linewidth}
      \centering
      \includegraphics[width=\linewidth]{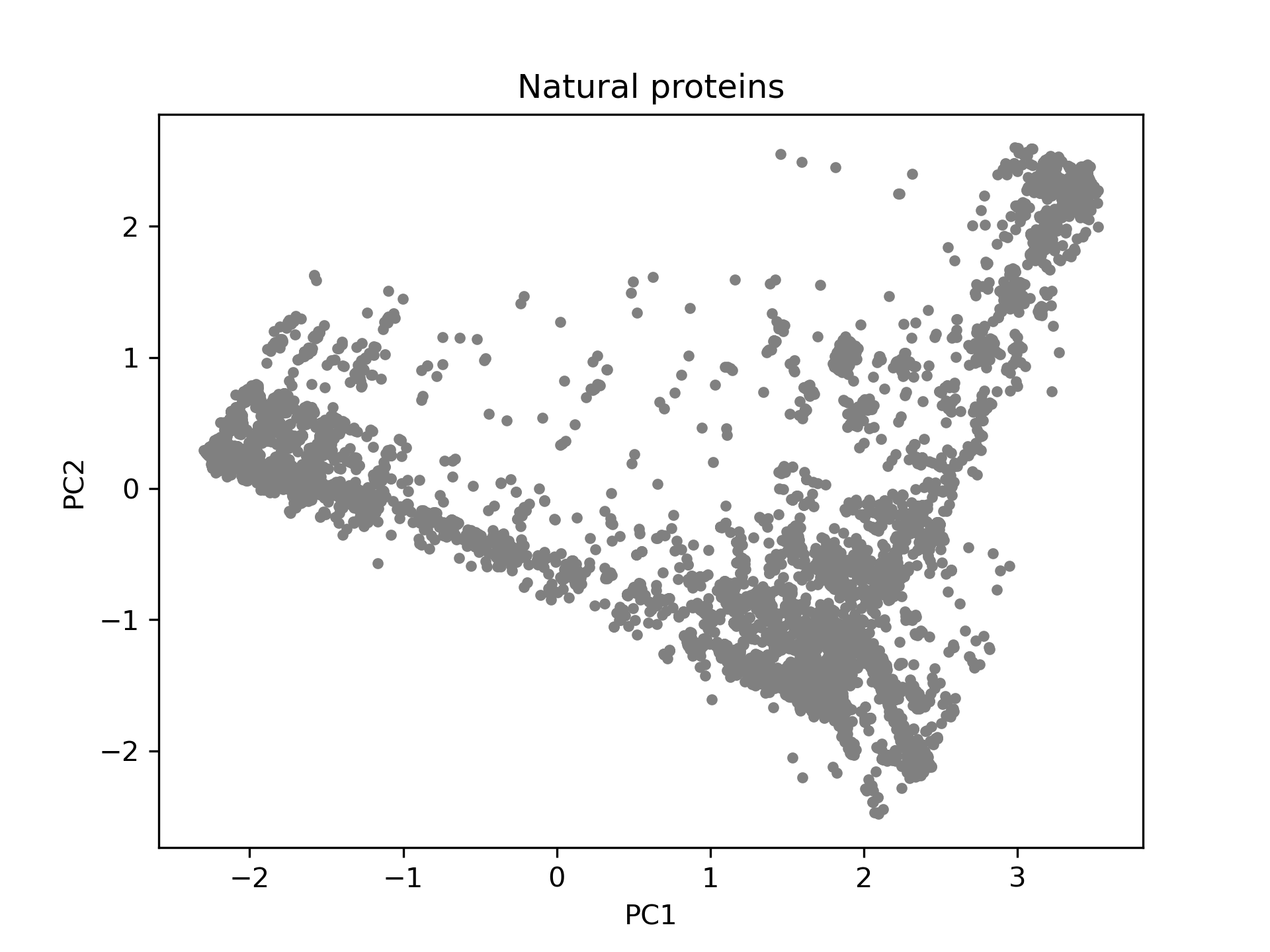}
      (a) Natural Proteins.
      \label{fig:vae_as}
    \end{minipage}
    \begin{minipage}[t]{0.4\linewidth}
      \centering
      \includegraphics[width=\linewidth]{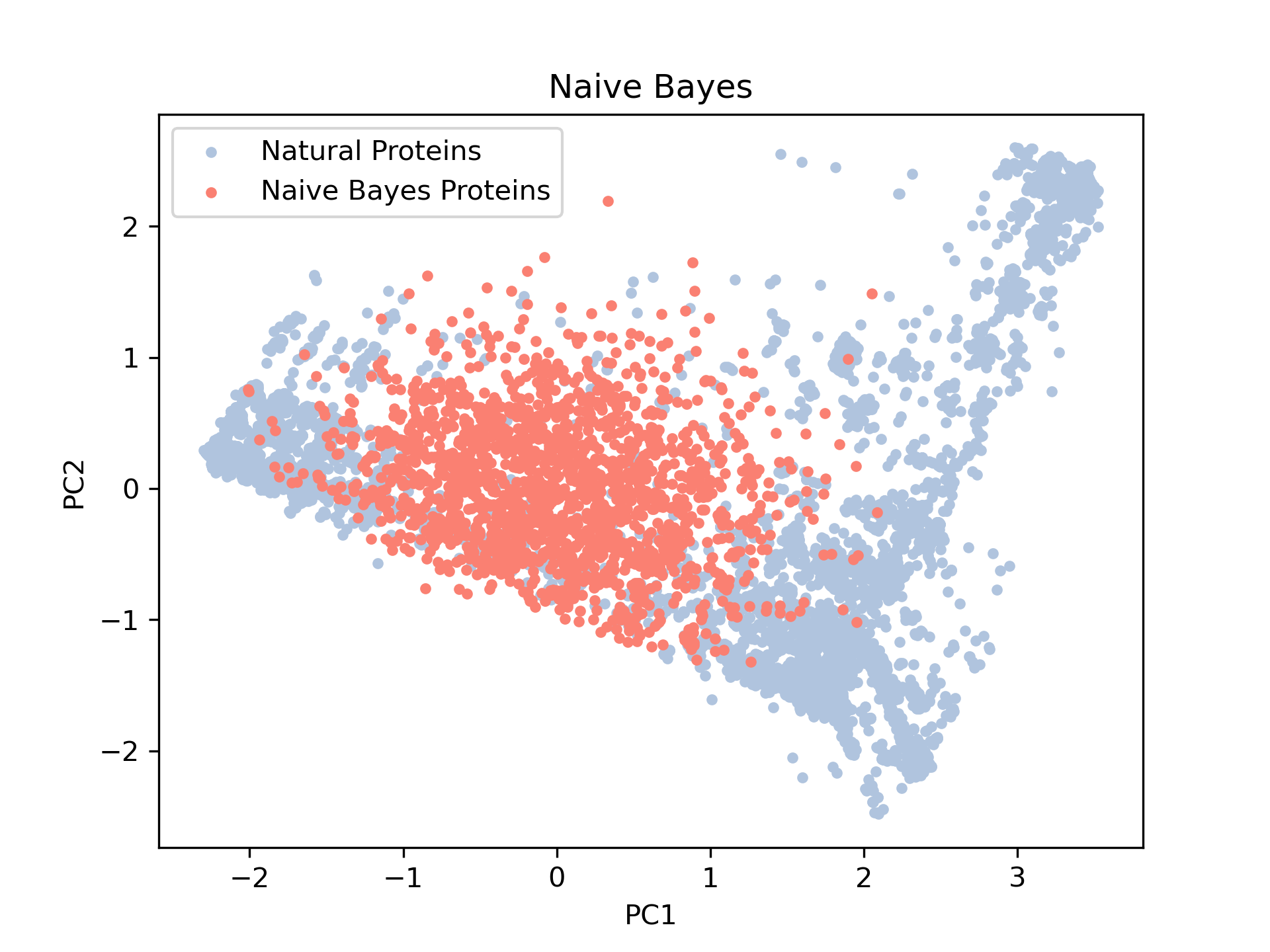}
      (b) Naive Bayes.
      \label{fig:lstm_as}
    \end{minipage}
    \begin{minipage}[t]{0.4\linewidth}
      \centering
      \includegraphics[width=\linewidth]{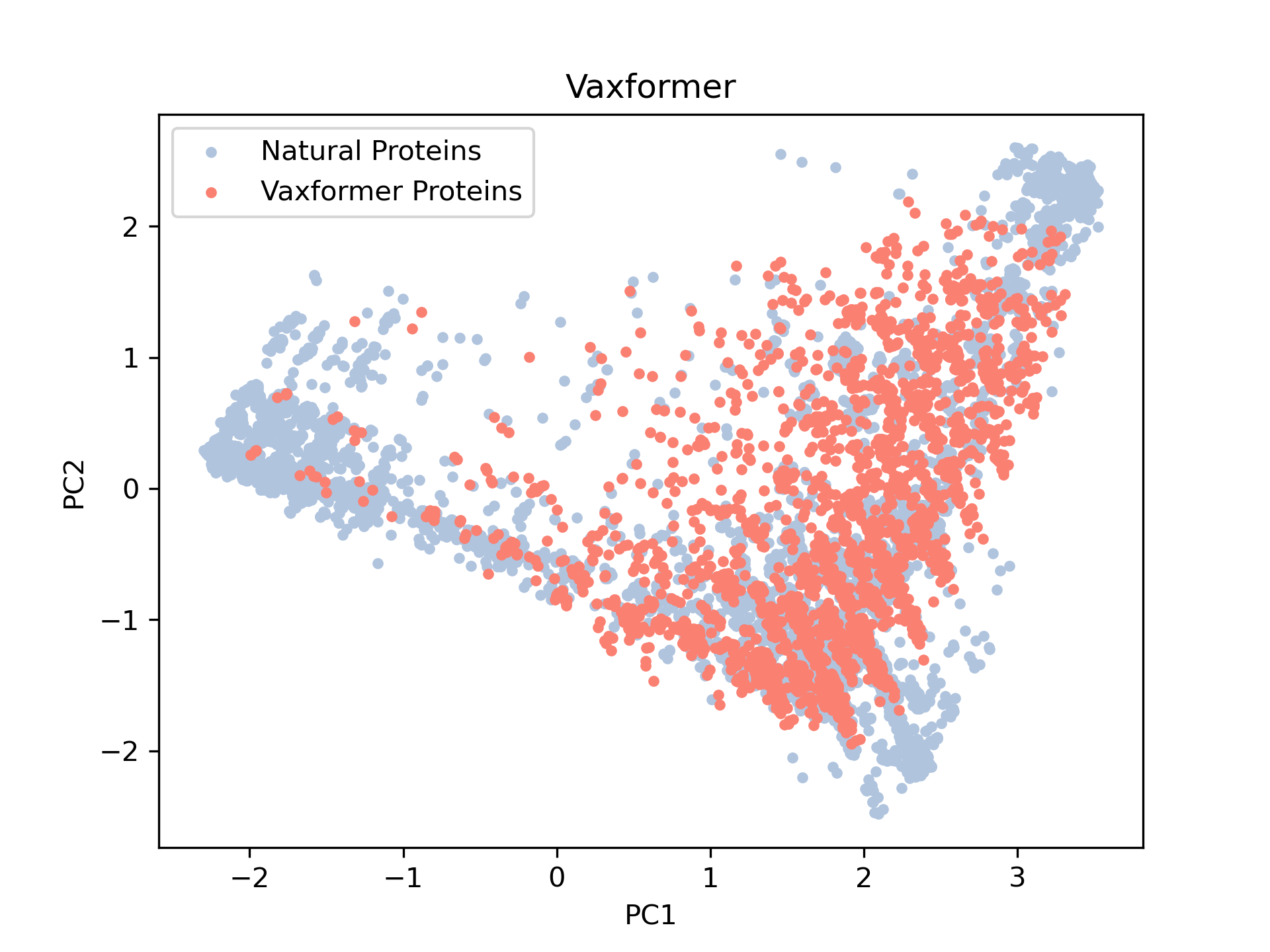}
      (c) Vaxformer.
      \label{fig:vaxformer_as}
    \end{minipage}
    \begin{minipage}[t]{0.4\linewidth}
      \centering
      \includegraphics[width=\linewidth]{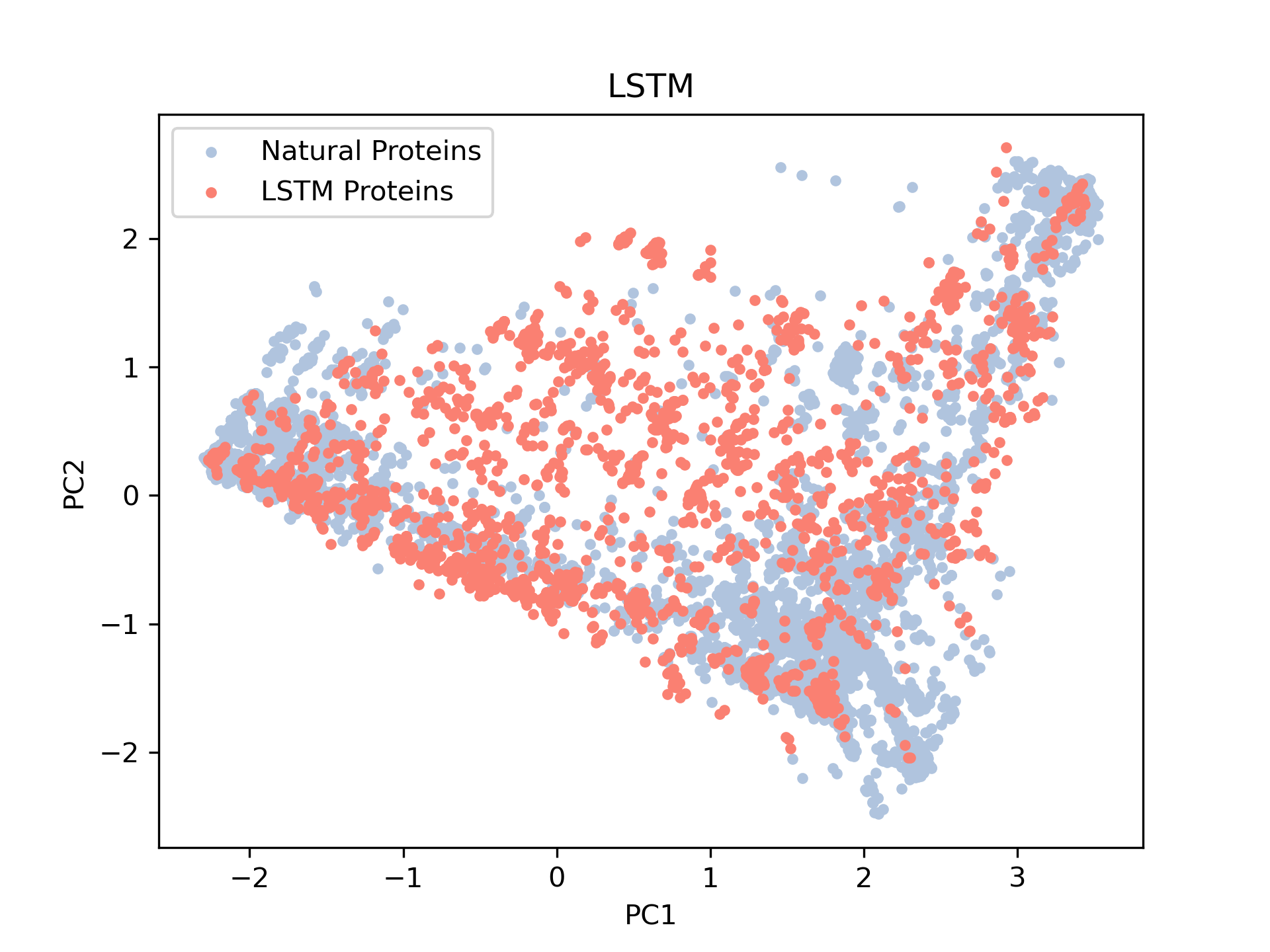}
      (d) LSTM.
      \label{fig:vaxformer_as}
    \end{minipage}
    \caption{PCA visualisation of the natural and generated sequences from the generative models: Naive Bayes, Vaxformer, and LSTM. Sequences from the training data are displayed in grey, while generated sequences are in orange. Each generative model covers different regions of the natural distribution. The baseline model (Naive Bayes) covers an area where natural proteins rarely occur, suggesting a bad fit, while LSTM and Vaxformer cover more favourable regions.}
    \label{fig:pca}
\end{figure*}

We also visualise the PCA distribution of the generated proteins to qualitatively verify that the generated proteins are relatively close to the natural proteins, which is displayed in Figure \ref{fig:pca}. The training data (natural proteins) were aligned using MSA and each of the generative models predict proteins following this alignment. All of the aligned proteins were then one-hot encoded and PCA transformation was fit on the training data (grey dots in Figure \ref{fig:pca}). The results of generative models were then transformed using the PCA projection (plotted in orange).

Note that each of the generative models covered different regions of the natural distribution. This suggests that by using diverse models, we are able to generate a wider range of proteins than using a single model. The Naive Bayes generates proteins lying in the region where natural proteins rarely occur, while the Vaxformer and LSTM produce sequences which are more likely to occur naturally. This provides additional validation that Vaxformer and LSTM perform better than the baseline.

\section{Conclusions}
\label{sec:concl}

In this study, we proposed Vaxformer, a novel conditional spike protein LM architecture capable of producing antigenicity-controlled SARS-CoV-2 spike proteins.
Vaxformer-generated sequences are stable and of valid structure, distributed along the principal variance directions of natural sequences, and well-modulated by their antigenicity scores.
Therefore, we argue that the Vaxformer-generated sequences can make a noteworthy contribution to the study of vaccine design against SARS-CoV-2.

We note that it is relatively trivial for the proposed LMs to reach the minima of respective loss functions, as shown by their respective perplexity scores.
However, it is important to note that a good performance in model evaluation metrics such as perplexity may not be directly associated with the model performance from a biological standpoint.
To address this issue, a proxy loss function can be created to emulate the biological evaluations and further enhance the ability of a model to generate biologically-sound sequences.

One possible solution that we have experimented with is to use a Generative Adversarial Network (GAN) architecture~\cite{goodfellowGenerativeAdversarialNets2014}, where the discriminator is trained to distinguish between real and fake sequences, and at the same time, to predict their antigenicity scores.
However, implementing GAN to discrete data such as AA sequences is very challenging since they are not differentiable.
One approach to solve this is by using SeqGAN~\cite{yuSeqGANSequenceGenerative2017} which employs policy gradient reinforcement learning~\cite{williamsSimpleStatisticalGradientfollowing1992}.
Unfortunately, implementing multiple Monte Carlo rollouts for 1,299 AAs in every training step can be very time-consuming.
Nonetheless, having a more efficient approach to adding a biologically-aware discriminator to evaluate the generated sequences during training can improve the model's biological performance.

Another potential avenue for exploration is to understand better the antigenicity information we get from netMHCpan evaluation. We could derive new metrics from the peptides hits number like the proportion of the protein that is exposed to the immune system, or detect specific segments of the protein that contributes more to the antigenicity score. By pinpointing which parts of a protein sequence contribute most significantly to its antigenicity, we may be able to improve the quality of the generated proteins and gain a more nuanced understanding of the factors that contribute to immune recognition.

In addition, future work could explore the possibility of increasing the granularity of the antigenicity score. While this study focused on three quantiles, there are no theoretical limitations preventing a more extended exploration. For example generating proteins with even higher scores than the natural ones would be of great interest for vaccine design. Such an investigation may further refine our understanding of the underlying characteristic differences of sequences with different antigenicity scores.

For future work, we also recommend using other tools than DDGun for assessing stability such as Rosetta \cite{barlow2018flex} or FoldX \cite{buss2018foldx} to validate results. Furthermore, the spike protein is composed of several domains which fold separately \cite{huang2020structural}. Perhaps this independency can be exploited by some hierarchical generative models or evaluation techniques such as Alphafold Multimer \cite{evans2021protein}. Furthermore, generating only one domain may allow using more computationally expensive models which were intractable in the case of the whole protein.

To support further innovation, the code used in this study is accessible via \href{https://github.com/aryopg/vaxformer}{https://github.com/aryopg/vaxformer}.





\section{Competing interests}
No competing interest is declared.

\section{Author contributions statement}

A.P.G. conceived the general experiments planning and worked on the training of protein language models.
M.K. worked on the DDGun and RMSE evaluation.
A.F. worked on the netMHCpan evaluation.
A.P.G., M.K., and A.F. analysed and interpreted the results jointly.
A.P.G., M.K. wrote the technical aspect of the manuscript.
A.F. wrote the biological background of the manuscript.
A.R., D.A.O, and J.A.A. reviewed the manuscript.

\section{Acknowledgments}
This work was supported by the United Kingdom Research and Innovation (grant EP/S02431X/1), UKRI Centre for Doctoral Training in Biomedical AI at the University of Edinburgh, School of Informatics. For the purpose of open access, the author has applied a creative commons attribution (CC BY) licence to any author accepted manuscript version arising.

This work was performed using resources provided by the Cambridge Service for Data Driven Discovery (CSD3) operated by the University of Cambridge Research Computing Service (www.csd3.cam.ac.uk), provided by Dell EMC and Intel using Tier-2 funding from the Engineering and Physical Sciences Research Council (capital grant EP/T022159/1), and DiRAC funding from the Science and Technology Facilities Council (www.dirac.ac.uk). We thank the PL-Grid and CI-TASK Infrastructure, Poland, for providing their hardware and software resources.

\bibliographystyle{plain}
\bibliography{main}

\begin{thebibliography}{10}

\bibitem{noauthor_covid-19_nodate}
{COVID}-19 vaccine tracker.

\bibitem{bairochUniversalProteinResource2004}
A.~Bairoch.
\newblock The {{Universal Protein Resource}} ({{UniProt}}).
\newblock {\em Nucleic Acids Research}, 33(Database issue):D154--D159, December
  2004.

\bibitem{barlow2018flex}
Kyle~A Barlow, Shane O~Conchuir, Samuel Thompson, Pooja Suresh, James~E Lucas,
  Markus Heinonen, and Tanja Kortemme.
\newblock Flex ddg: Rosetta ensemble-based estimation of changes in
  protein--protein binding affinity upon mutation.
\newblock {\em The Journal of Physical Chemistry B}, 122(21):5389--5399, 2018.

\bibitem{bishop2006pattern}
Christopher~M Bishop and Nasser~M Nasrabadi.
\newblock {\em Pattern recognition and machine learning}, volume~4.
\newblock Springer, 2006.

\bibitem{buss2018foldx}
Oliver Bu{\ss}, Jens Rudat, and Katrin Ochsenreither.
\newblock Foldx as protein engineering tool: better than random based
  approaches?
\newblock {\em Computational and structural biotechnology journal}, 16:25--33,
  2018.

\bibitem{delano2002pymol}
Warren~L DeLano et~al.
\newblock Pymol: An open-source molecular graphics tool.
\newblock {\em CCP4 Newsl. Protein Crystallogr}, 40(1):82--92, 2002.

\bibitem{dolginTangledHistoryMRNA2021}
Elie Dolgin.
\newblock The tangled history of mrna vaccines.
\newblock {\em Nature}, 597(7876):318--324, September 2021.

\bibitem{dosovitskiy2016generating}
Alexey Dosovitskiy and Thomas Brox.
\newblock Generating images with perceptual similarity metrics based on deep
  networks.
\newblock {\em Advances in neural information processing systems}, 29, 2016.

\bibitem{edgar2004muscle}
Robert~C Edgar.
\newblock Muscle: a multiple sequence alignment method with reduced time and
  space complexity.
\newblock {\em BMC bioinformatics}, 5(1):1--19, 2004.

\bibitem{evans2021protein}
Richard Evans, Michael O’Neill, Alexander Pritzel, Natasha Antropova, Andrew
  Senior, Tim Green, Augustin {\v{Z}}{\'\i}dek, Russ Bates, Sam Blackwell,
  Jason Yim, et~al.
\newblock Protein complex prediction with alphafold-multimer.
\newblock {\em BioRxiv}, pages 2021--10, 2021.

\bibitem{federhenNCBITaxonomyDatabase2012}
S.~Federhen.
\newblock The {{NCBI Taxonomy}} database.
\newblock {\em Nucleic Acids Research}, 40(D1):D136--D143, January 2012.

\bibitem{ferruzProtGPT2DeepUnsupervised2022}
Noelia Ferruz, Steffen Schmidt, and Birte H{\"o}cker.
\newblock {{ProtGPT2}} is a deep unsupervised language model for protein
  design.
\newblock {\em Nature Communications}, 13(1):4348, July 2022.

\bibitem{finnPfamProteinFamilies2014}
Robert~D. Finn, Alex Bateman, Jody Clements, Penelope Coggill, Ruth~Y.
  Eberhardt, Sean~R. Eddy, Andreas Heger, Kirstie Hetherington, Liisa Holm,
  Jaina Mistry, Erik L.~L. Sonnhammer, John Tate, and Marco Punta.
\newblock Pfam: The protein families database.
\newblock {\em Nucleic Acids Research}, 42(D1):D222--D230, January 2014.

\bibitem{flemming_omicron_2022}
Alexandra Flemming.
\newblock Omicron, the great escape artist.
\newblock 22(2):75--75, 2022.
\newblock Number: 2 Publisher: Nature Publishing Group.

\bibitem{goodfellowGenerativeAdversarialNets2014}
Ian Goodfellow, Jean {Pouget-Abadie}, Mehdi Mirza, Bing Xu, David
  {Warde-Farley}, Sherjil Ozair, Aaron Courville, and Yoshua Bengio.
\newblock Generative {{Adversarial Nets}}.
\newblock In {\em Advances in {{Neural Information Processing Systems}}},
  volume~27. {Curran Associates, Inc.}, 2014.

\bibitem{gromiha2010protein}
M~Michael Gromiha.
\newblock {\em Protein bioinformatics: from sequence to function}.
\newblock academic press, 2010.

\bibitem{hochreiterLongShortTermMemory1997}
Sepp Hochreiter and J{\"u}rgen Schmidhuber.
\newblock Long {{Short-Term Memory}}.
\newblock {\em Neural Computation}, 9(8):1735--1780, November 1997.

\bibitem{huangStructuralFunctionalProperties2020}
Yuan Huang, Chan Yang, Xin-feng Xu, Wei Xu, and Shu-wen Liu.
\newblock Structural and functional properties of {{SARS-CoV-2}} spike protein:
  Potential antivirus drug development for {{COVID-19}}.
\newblock {\em Acta Pharmacologica Sinica}, 41(9):1141--1149, September 2020.

\bibitem{huang2020structural}
Yuan Huang, Chan Yang, Xin-feng Xu, Wei Xu, and Shu-wen Liu.
\newblock Structural and functional properties of sars-cov-2 spike protein:
  potential antivirus drug development for covid-19.
\newblock {\em Acta Pharmacologica Sinica}, 41(9):1141--1149, 2020.

\bibitem{jumper2021highly}
John Jumper, Richard Evans, Alexander Pritzel, Tim Green, Michael Figurnov,
  Olaf Ronneberger, Kathryn Tunyasuvunakool, Russ Bates, Augustin
  {\v{Z}}{\'\i}dek, Anna Potapenko, et~al.
\newblock Highly accurate protein structure prediction with alphafold.
\newblock {\em Nature}, 596(7873):583--589, 2021.

\bibitem{keskarCTRLConditionalTransformer2019}
Nitish~Shirish Keskar, Bryan McCann, Lav~R. Varshney, Caiming Xiong, and
  Richard Socher.
\newblock {{CTRL}}: {{A Conditional Transformer Language Model}} for
  {{Controllable Generation}}, September 2019.

\bibitem{khare2021gisaid}
Shruti Khare, C{\'e}line Gurry, Lucas Freitas, Mark~B Schultz, Gunter Bach,
  Amadou Diallo, Nancy Akite, Joses Ho, Raphael~TC Lee, Winston Yeo, et~al.
\newblock Gisaid’s role in pandemic response.
\newblock {\em China CDC weekly}, 3(49):1049, 2021.

\bibitem{kingmaAdamMethodStochastic2017}
Diederik~P. Kingma and Jimmy Ba.
\newblock Adam: {{A Method}} for {{Stochastic Optimization}}, January 2017.

\bibitem{larsen2016autoencoding}
Anders Boesen~Lindbo Larsen, S{\o}ren~Kaae S{\o}nderby, Hugo Larochelle, and
  Ole Winther.
\newblock Autoencoding beyond pixels using a learned similarity metric.
\newblock In {\em International conference on machine learning}, pages
  1558--1566. PMLR, 2016.

\bibitem{leinonenUniProtArchive2004}
Rasko Leinonen, Federico~Garcia Diez, David Binns, Wolfgang Fleischmann,
  Rodrigo Lopez, and Rolf Apweiler.
\newblock {{UniProt}} archive.
\newblock {\em Bioinformatics}, 20(17):3236--3237, November 2004.

\bibitem{madaniLargeLanguageModels2023}
Ali Madani, Ben Krause, Eric~R. Greene, Subu Subramanian, Benjamin~P. Mohr,
  James~M. Holton, Jose~Luis Olmos, Caiming Xiong, Zachary~Z. Sun, Richard
  Socher, James~S. Fraser, and Nikhil Naik.
\newblock Large language models generate functional protein sequences across
  diverse families.
\newblock {\em Nature Biotechnology}, January 2023.

\bibitem{montanucci2019ddgun}
Ludovica Montanucci, Emidio Capriotti, Yotam Frank, Nir Ben-Tal, and Piero
  Fariselli.
\newblock Ddgun: an untrained method for the prediction of protein stability
  changes upon single and multiple point variations.
\newblock {\em BMC bioinformatics}, 20:1--10, 2019.

\bibitem{phillipsGeneratingImmuneawareSARSCoV22022}
Dominic Phillips, Hans-Christof Gasser, Sebestyen Kamp, Aleksander
  Pa{\l}kowski, and Lukasz Rabalski.
\newblock Generating {{Immune-aware SARS-CoV-2 Spike Proteins}} for {{Universal
  Vaccine Design}}.
\newblock 2022.

\bibitem{radfordLanguageModelsAre2019}
Alec Radford, Jeffrey Wu, Rewon Child, David Luan, Dario Amodei, and Ilya
  Sutskever.
\newblock Language {{Models}} are {{Unsupervised Multitask Learners}}.
\newblock February 2019.

\bibitem{reynissonNetMHCpan4NetMHCIIpan4Improved2020}
Birkir Reynisson, Bruno Alvarez, Sinu Paul, Bjoern Peters, and Morten Nielsen.
\newblock {{NetMHCpan-4}}.1 and {{NetMHCIIpan-4}}.0: Improved predictions of
  {{MHC}} antigen presentation by concurrent motif deconvolution and
  integration of {{MS MHC}} eluted ligand data.
\newblock {\em Nucleic Acids Research}, 48(W1):W449--W454, July 2020.

\bibitem{shakhnovichImmunogenicityClinicalPractice2020}
Valentina Shakhnovich, Bernd Meibohm, Amy Rosenberg, Andrzej~M. Kierzek, Rachel
  Hasenkamp, Ryan~S. Funk, Craig~J. Thalhauser, Piet~H. {van der Graaf},
  Yow-Ming~C. Wang, and Lora Hamuro.
\newblock Immunogenicity in {{Clinical Practice}} and {{Drug Development}}:
  {{When}} is it {{Significant}}?
\newblock {\em Clinical and Translational Science}, 13(2):219--223, March 2020.

\bibitem{shinProteinDesignVariant2021}
Jung-Eun Shin, Adam~J. Riesselman, Aaron~W. Kollasch, Conor McMahon, Elana
  Simon, Chris Sander, Aashish Manglik, Andrew~C. Kruse, and Debora~S. Marks.
\newblock Protein design and variant prediction using autoregressive generative
  models.
\newblock {\em Nature Communications}, 12(1):2403, April 2021.

\bibitem{shuGISAIDGlobalInitiative2017}
Yuelong Shu and John McCauley.
\newblock {{GISAID}}: {{Global}} initiative on sharing all influenza data
  \textendash{} from vision to reality.
\newblock {\em Eurosurveillance}, 22(13):30494, March 2017.

\bibitem{vaswaniAttentionAllYou2017}
Ashish Vaswani, Noam Shazeer, Niki Parmar, Jakob Uszkoreit, Llion Jones,
  Aidan~N. Gomez, Lukasz Kaiser, and Illia Polosukhin.
\newblock Attention {{Is All You Need}}, December 2017.

\bibitem{wang_novel_2020}
Chen Wang, Peter~W. Horby, Frederick~G. Hayden, and George~F. Gao.
\newblock A novel coronavirus outbreak of global health concern.
\newblock 395(10223):470--473, 2020.
\newblock Publisher: Elsevier.

\bibitem{williamsSimpleStatisticalGradientfollowing1992}
Ronald~J. Williams.
\newblock Simple statistical gradient-following algorithms for connectionist
  reinforcement learning.
\newblock {\em Machine Learning}, 8(3):229--256, May 1992.

\bibitem{yuSeqGANSequenceGenerative2017}
Lantao Yu, Weinan Zhang, Jun Wang, and Yong Yu.
\newblock {{SeqGAN}}: {{Sequence Generative Adversarial Nets}} with {{Policy
  Gradient}}, August 2017.

\end{thebibliography}



\end{document}